\documentclass[a4paper,sort&compress,5p]{elsarticle}
\usepackage{amsmath}
\usepackage{amsfonts}
\usepackage{amssymb}
\usepackage{array}
\usepackage{multirow}
\usepackage{booktabs}
\usepackage{hyperref}
\usepackage[all]{hypcap}
\usepackage{graphicx} % Include figure files
\usepackage{dcolumn} % Align table columns on decimal point
\usepackage{paralist}

\journal{Physics Letters B}

\newif\ifonecol\onecolfalse

\begin{document}
\title{Diffuse Ultra-High Energy Neutrino Fluxes and Physics Beyond the Standard Model}
\date{\today}

%%---------------AUTHORS--------------------%%
\author[rvt]{Atri Bhattacharya}
%%\email{atri@hri.res.in}
%%\address{Harish-Chandra Research Institute, Chhatnag Road, Jhunsi, Allahabad 211 019, India}

\author[rvt]{Sandhya Choubey}
%%\email{sandhya@hri.res.in}
%%\address{Harish-Chandra Research Institute, Chhatnag Road, Jhunsi, Allahabad 211 019, India}

\author[rvt]{Raj Gandhi}
%%\email{nubarnu@gmail.com}
%%\address{Harish-Chandra Research Institute, Chhatnag Road, Jhunsi,Allahabad 211 019, India}

\author[els]{Atsushi Watanabe}
%%\email{atsushi@hri.res.in}
\address[rvt]{Harish-Chandra Research Institute, Chhatnag Road, Jhunsi,Allahabad 211 019, India}
\address[els]{Department of Physics, Kyushu University, Fukuoka 812-8581, Japan}

%%------------------------------------------%%

%%-----------------ABSTRACT------------------%%
\begin{abstract}
We study spectral distortions of diffuse ultra-high energy (UHE) neutrino flavour fluxes resulting due to physics beyond the Standard Model (SM). Even large spectral differences between flavours at the source are massaged into a common shape at earth by SM oscillations, thus, any significant observed spectral differences are an indicator of new physics present in the oscillation probability during propagation. Lorentz symmetry violation (LV) and Neutrino decay are examples, and result in significant distortion of the fluxes and of the well-known bounds on them, which may allow UHE detectors to probe LV parameters, lifetimes and the mass hierarchy over a broad range. 
\end{abstract}
%%-------------------------------------------%%
%\pacs{14.60.Pq,14.60.Lm,14.60.St,13.15.+g,11.30.Cp,98.54.Cm,95.85.Ry}
%{11.30.Er.,11.30.Cp.,14.60.Pq,13.15.+g}
%{11.30.Er,11.30.Pb,12.60.Jv}
%11.30.Cp Lorentz and Poincare invariance 
%14.60.Pq Neutrino mass and mixing
%14.60.St Non-standard-model neutrinos, right-handed neutrinos, etc
%14.60.Lm Ordinary Neutrinos
%%%%%%%%%%%%%%%%%%%%%%%%%%%%%%%%%%%%%%%%%%%%%%%%%%%%%%%%%%%%%%%%%%%%%%%%%%

\maketitle

\section{\label{sec:intro}Introduction}
The neutrino sky spans about twenty five orders of magnitude in energy, potentially offering the possibility of probing the universe at widely disparate
energy scales. The high end ($10^{11}-10^{12} $GeV) of this remarkably broad band in energy is set by: a) GZK neutrinos \cite{PhysRevLett.16.748,Zatsepin:1966jv}, which originate in the interactions of the highest energy cosmic rays with the cosmic microwave photon background and b) neutrinos from the most energetic astrophysical objects observed in the universe, \textit{i.e.} active galactic nuclei (AGNs) and Gamma-Ray bursts (GRBs). Detection, still in the future, presents both considerable opportunities and formidable challenges. In particular, at the very highest energies ($10^5$ GeV and above) which are the focus of this paper, the tiny fluxes that arrive at earth require detectors that combine the capability to monitor very large detection volumes with innovative techniques (for reviews, see e.g. \cite{Hoffman:2008yu} and \cite{Halzen:2007sz}). Examples of such detectors are AMANDA \cite{Baret:2008zz}, ICECUBE \cite{Halzen:2009tz}, BAIKAL \cite{Aynutdinov:2009zz}, ANTARES \cite{Margiotta:2009zz}, RICE \cite{1742-6596-81-1-012008} and ANITA \cite{1742-6596-136-2-022052}.

A compelling motivation for exploring UHE  neutrino astronomy is the fact that the origin of cosmic rays (CR) beyond the ``knee" ($10^6$ GeV) remains a mystery many decades after their discovery. Additionally, CR with energies in excess of $10^{11}$ GeV have been observed\cite{Roth:2009zz,Abbasi:2007sv}, signalling the presence of astrophysical particle accelerators of unprecedentedly high energies. If protons as well as neutrons are accelerated at these sites in addition to electrons, standard particle physics predicts correlated fluxes of neutrons and neutrinos which escape from the confining magnetic field of the source, while protons and electrons stay trapped. Generically, the electrons lose energy rapidly via synchrotron radiation. These radiated photons provide a target for the accelerated protons, which results in the production of pions, muons and ultimately, neutrinos in the ratio ${\nu_e:\nu_\mu:\nu_\tau = 1:2:0}$. The detection and study of ultra-high energy (UHE) neutrinos is thus a probe of the origin of CR and the physics of UHE astrophysical accelerators.

As is well understood due to a wealth of data from solar, atmospheric, reactor and accelerator  experiments, neutrinos have mass and oscillate into one another. These  experiments have led to determinations, to various degrees of accuracy, of the neutrino mass (squared) differences $ \Delta m^2_{ij} $ between mass eigenstates $ i $ and $ j $, and the mixing angles $ U_{\alpha i} $ \cite{Schwetz:2009zz}. The latter characterize the overlap of neutrino mass (or propagation) eigenstates (denoted by $\nu_i, i=1,2,3$) and the interaction eigenstates (denoted by $\nu
_\alpha, \alpha=e,\mu,\tau$). Over the very large distances traversed by neutrinos from 
the most energetic extragalactic hadronic accelerators, the initial source ratios ${\nu_e^s:\nu_\mu^s:\nu_\tau^s = 1:2:0}$
will, due to (vacuum) oscillations, transmute, by the time they reach a terrestrial detector, into $\nu_e^d:\nu_\mu^d:\nu_\tau^d = 1:1:1$ \cite{Learned:1994wg,Athar:2000yw}.

A series of papers \cite{Beacom:2002vi,Beacom:2003nh,Barenboim:2003jm,Keranen:2003xd,Beacom:2003eu,Hooper:2004xr,Hooper:2005jp, Meloni:2006gv,Xing:2008fg,Esmaili:2009dz} over the past few years have demonstrated that if the flavour ratios $\nu_e^d:\nu_\mu^d:\nu_\tau^d$ detected by extant and upcoming neutrino telescopes were to deviate significantly from this democratic prediction, then important conclusions about physics beyond the Standard Model and neutrino oscillation parameters may consequently be inferred. In addition, deviations of these  measured ratios have been shown to be sensitive to  neutrino oscillation parameters \cite{Farzan:2002ct, Beacom:2003zg, Serpico:2005sz, Xing:2006xd, Rodejohann:2006qq, Winter:2006ce, Awasthi:2007az, Lipari:2007su, Blum:2007ie, Pakvasa:2007dc, Choubey:2008di, Esmaili:2009dz} (\textit{e.g.}~the  mixing angles  and the Dirac CP violating phase).

In this paper we  study the spectral distortions in the {\it diffuse} ({\it i.e.} integrated over source distribution and redshift) UHE neutrino flux as a probe for the  effects of new physics. For specificity, we focus on AGN fluxes, and use, as a convenient bench-mark, the well-known upper bounds first derived by Waxman and Bahcall (WB) \cite{Waxman:1998yy} and later by Mannheim, Protheroe and Rachen (MPR) \cite{Mannheim:1998wp} on such fluxes for both neutron-transparent  and neutron-opaque sources (or, equivalently, sources that are \emph{optically thin} and \emph{optically thick}, respectively, to the emission of neutrons). In particular we focus on the upper bounds to the diffuse neutrino flux from hadronic photoproduction in AGN's derived in \cite{Mannheim:1998wp} using the experimental upper limit on cosmic ray protons. All distortions in the fluxes are, as would be expected, transmitted to the upper bounds, thus providing a convenient way of representing and studying them.

Prior to this, we first demonstrate (in Sec.~\ref{sec:spectral-averaging}) that the usual (SM) neutrino oscillations not only tend to equilibrate widely differing source flux magnitudes between flavours, but also massage them into a common spectral shape, as one would intuitively expect. Thus observed relative spectral distortions among flavours are a probe  of new physics present in the propagation equation. To demonstrate our approach we then  focus on two specific examples, \textit{i.e.,}
\begin{inparaenum}[\itshape a\upshape)]
\item Lorentz violation (LV), and
\item Incomplete decay of neutrinos
\end{inparaenum}
 in the neutrino sector. Our method can straightforwardly be applied to other new physics scenarios and our results translated into bounds on muon track versus shower event rates\footnote{ These count the sum of 
\begin{inparaenum}[\itshape a\upshape)]
\item neutral current (NC) events of all flavours, 
\item electron neutrino charged current (CC) events at all energies, and
\item $\nu_\tau$ induced CC events at energies below $\leq$ 1 PeV ($10^6$ GeV),
\end{inparaenum}
 whereas muon track events arise from $\nu_\mu$ induced muons born in  
CC interactions.} for UHE experiments.

In our calculations we use the current best-fit values of neutrino mixing paramenters as given in \cite{Maltoni:2008ka}. The mass squared differences and mixing angles are
\begin{gather*}
	\begin{split}
	\Delta m_{21}^2 &= 7.65 \times 10^{-5} \text{ eV}^2 \\
	\Delta m_{31}^2 &= \pm 2.40 \times 10^{-3} \text{ eV}^2
	\end{split}
	\\
	\sin^2(\theta_{12})=0.321,\;\; \sin^2(\theta_{23})=0.47,\;\; \sin^2(\theta_{13})=0.003.
\end{gather*}

\section{\label{sec:base-flux}AGN Fluxes and Bounds} Physics models of hadronic blazars, while they may differ in their details, tightly co-relate emitted luminosities of cosmic rays, gamma rays and neutrinos from such sources (\cite{Mannheim:1998wp}). The standard approach to calculating them consists in using an astrophysical model to determine the spectral shape and then normalizing the flux using the extra-galactic gamma-ray background (EGRB) (\cite{Mannheim:1998wp}) and the observed cosmic-ray background \cite{Waxman:1998yy,Mannheim:1998wp}.

We follow this procedure here in order to  calculate the base flux \textit{i.e.}, that without the effects of decay or Lorentz-violation, for the neutron-opaque sources. First, the generic neutrino production spectrum from a single source is approximated by
\begin{equation}
	Q_n(E_n,L_p) \propto L_p \exp\left[\frac{-E_n}{E_{\rm max}}\right]
	\left \{ \begin{array}{ll}
	        E_n^{-1}E_b^{-1} & (E_n < E_b)\\
                E_n^{-2} & (E_b < E_n)
                \end{array} \right. .
        \label{eq.generic_neutron}
\end{equation}
In the above equation, $ L_p $ represents the proton luminosity of the source, $ E_{max} $ represents the cut-off energy beyond which the spectrum rapidly falls off, $ E_b $ represents the energy at which the spectrum changes shape (specifically from a rising dependence on energy below $ E_b $ to a plateau beyond it, in terms of $ E_n^2 Q_n $), and $ Q_n $ and $ E_n $ represent the neutron spectrum and energy respectively. 

To relate the neutrino emission spectra to the neutron spectra given by Eq.~\eqref{eq.generic_neutron} it is useful to express the energies of each of the secondary particles produced as a fraction $ \zeta $ of the proton energy transferred to it per interaction. For neutrinos, gamma-rays and neutrons, from AGN's
\begin{align}
	\xi_\nu \approx \xi_\gamma &\approx 0.1 \\
	\xi_n &\approx 0.5 .
	\label{eq.xi}
\end{align}
The energy per particle in units of the proton energy for neutrinos and neutrons are found to be $ \langle E_\nu \rangle /E_p \approx 0.033 $ and $ \langle E_n \rangle /E_p \approx 0.83 $ respectively, giving the relative energy of neutrinos and neutrons escaping from the source as
\begin{equation}
	\eta_{\nu n} = \langle E_\nu \rangle / \langle E_n \rangle \approx 0.04 
	\label{eq.eta_nu-n}
\end{equation}
With the $ Q_{n} $ known, the neutrino spectrum at source can now be related to it using the quantities defined in Eq.~\eqref{eq.xi} and \eqref{eq.eta_nu-n}. Thus,
\begin{equation}
\begin{split}
Q_{\nu_\mu}(E) &= \frac{2}{3}\frac{\langle \xi_\nu \rangle}{\langle \xi_n \rangle \eta_{\nu n}^2}Q_n(E/\eta_{\nu n}) \\
&= 416Q_n(25E) .
\label{eq.generic_nu}
\end{split} 
\end{equation}
Since the flux of neutrinos is closely related to the emitted neutron flux as Eq.~\eqref{eq.generic_nu} shows, the diffuse neutrino flux due to sources that are optically thin to the emission of neutrons (or neutron-transparent sources) is considerably different from that due to sources which are optically thick (or neutron-opaque). When considering the modification of the diffuse neutrino flux bounds we use upper bounds to fluxes from both these kinds of photohadronic sources.

By varying the break energy $ E_b $ from $ 10^7 - 10^{10} $ GeV and maximally superposing these curves, we obtain the bound for the neutrino fluxes at source. Accounting for the red-shift in energy we replace $ E $ by $ E(1+z) $ in Eq.\eqref{eq.generic_nu}. Finally, we put in the effects of standard oscillations, and the total source flux is propagated from source to the earth by integrating over the red-shift factor $ z $. It is normalized as noted above. A related procedure is adopted for the optically thin curves.

The results of our calculation, when undistorted by non-standard physics, are presented as gray lines in both Fig.~\ref{lv1} and \ref{decay}, in the form of the WB and MPR bounds\footnote{Our (undistorted) bounds are similar, but not exactly the same as the MPR bounds, since we use a less elaborate propagation algorithm and neglect, in particular, photo-pion losses and secondary particle production, compared to the continuous-loss approximation used by MPR (\cite[Eq. 24, 26 and 19]{Mannheim:1998wp}). This does not in any significant way affect the conclusions drawn in our paper. Additionally, our bounds incorporate the effects of oscillations, which were not included by MPR.} (for both neutron transparent and neutron-opaque sources) along with present and projected sensitivities of various experiments. 

\section{\label{sec:spectral-averaging}Spectral Averaging due to Oscillations} 
\begin{figure}[htb]
	\centering
	\includegraphics[scale=0.27]{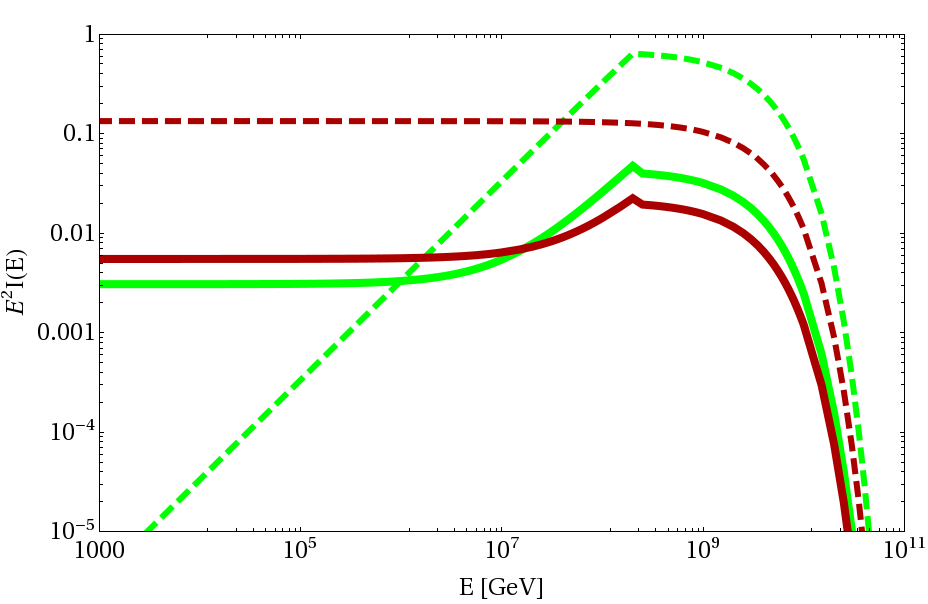}
	\caption{\label{osc-shape}The even-ing out of possible spectral distortions present at source due to oscillations over large distances. The dotted lines represent assumed spectra in a single source AGN for the flavours $ \nu_e $ (green) and $ \nu_\mu$ (deep red), the solid lines represent the corresponding diffuse fluxes at earth after integrating over source distribution and oscillations. In all of our figures, $I(E)$ denotes the diffuse flux spectrum of  flavours at earth, obtained as described in the text.}
\end{figure}

Fig.~\ref{osc-shape}  (in arbitrary units, and without normalisation) shows the spectra\footnote{The spectra shown here is unrealistic and chosen only to demonstrate the effect of standard oscilaltions on even such widely differing flavour fluxes.} of two flavours in a single source AGN, intentionally chosen to be significantly differing  in shape and magnitude, and the resulting diffuse fluxes from all such sources for the same flavours as seen at earth after standard propagation using the procedure described above. It is evident that not only do oscillations tend to bring widely differing  magnitudes close (to within a factor of 2) to each other, but they wash out even large differences in spectral shapes that may originate in a particular source, perhaps due to conventional physics, as \textit{e.g.} in \cite{PhysRevLett.95.181101}. We have checked that this conclusion holds in general, and  a common intermediate shape is assumed by both fluxes at earth detectors. These conclusions are  no longer true if in the propagation equation, the oscillation probability is modified by new physics in an energy-dependent manner, as we demonstrate in the examples below.

\section{Violations of Lorentz Symmetry} Violations of Lorentz symmetry (LV), if present, must be extremely tiny. They may appear in certain theories which are low energy limits of string theory \cite{Madore:2000en,Kostelecky:2003fs}, or could possibly signal the breakdown of the CPT theorem \cite{Greenberg:2002uu}. Additionally, if quantum gravity demands a fundamental length scale, leading to a breakdown of special relativity, or loop quantum gravity \cite{Rovelli:1994ge,Gambini:1998it,Alfaro:1999wd,Thiemann:2001yy,AmelinoCamelia:2003xp,Freidel:2003sp} leads to discrete space-time , one expects tiny LV effects to percolate to lower energies. For a recent discussion see \cite{Mattingly:2009jf} and references therein. UHE neutrinos, with their high energies and long oscillation baselines present a unique opportunity for testing these theories. Their effects in the context of flavour flux ratios have been discussed in \cite{Hooper:2005jp}. Here we demonstrate their effects on diffuse UHE fluxes (or equivalently, on the bounds thereon) by a representative calculation. For specificity we pick the low energy limit of string theory represented by the Standard model Extension \cite{Kostelecky:2003fs} and the corresponding modified dispersion relation implied by it. We consider, for simplicity, the two-flavour case with $\nu_\mu-\nu_\tau$ oscillations and a single real of{}f-diagonal Lorentz and CPT violating parameter $a$ with dimensions of mass, which modifies the effective hamiltonian (in the mass eigenstate basis) to 

\begin{equation}\label{LV_eqn1}
	H_{\mathrm{eff}}=\begin{pmatrix}
			\frac{m_1^2}{2E} & a \\
			a & \frac{m_2^2}{2E} \\
	\end{pmatrix} ,
\end{equation}
and gives an oscillation probability\footnote{For calculations in this section we use $ \theta_{23} = 45^o $ so that standard oscillation ensures complete symmetry between $ \nu_\mu $ and $ \nu_\tau $.}
\begin{equation}\label{probability-lv}
	P\left[ \nu_\mu \rightarrow \nu_\tau \right] = \frac{1}{4}\left( 1 - \frac{a^2}{\Omega^2} - \frac{\omega^2}{\Omega^2}\cos \left( 2\Omega L \right) \right)
\end{equation}
where $ \omega=\frac{\Delta m^2}{4E} $ and $ \Omega = \sqrt{\omega^2 + a^2} $.

\begin{figure}[ht]
\centering
\includegraphics[scale=0.28]{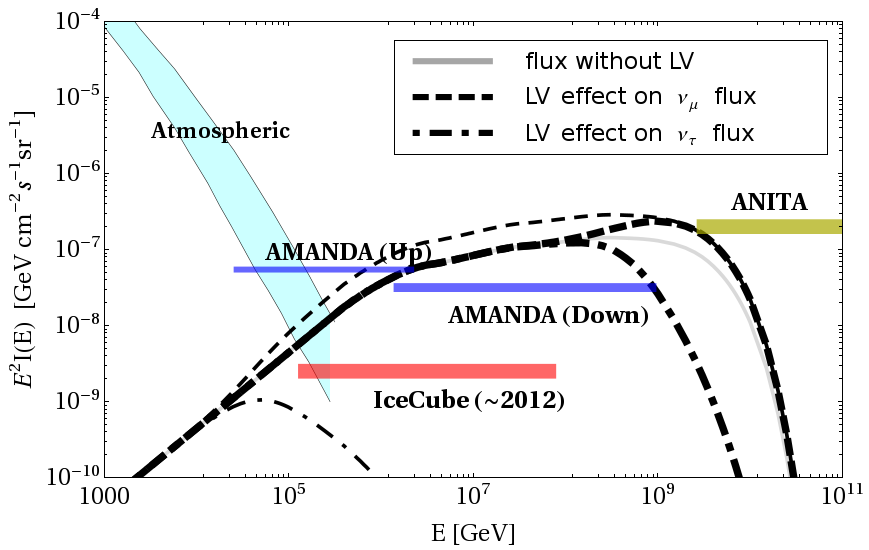}
\caption{\label{lv1}\textbf{Effect of Lorentz violation} on the $ \nu_{\mu}-\nu_{\tau} $ diffuse flux with lorentz violating parameter $a=10^{-26}$ GeV (thinner lines) and $ 10^{-30} $ GeV (in thick).}
\end{figure}

To calculate the diffuse fluxes for each of the two flavours we follow a procedure similar to that used to obtain the base flux in our plots (see Sec.~\ref{sec:base-flux}) but with the standard transition probability replaced by Eq.~\eqref{probability-lv}. The results are depicted in Fig.~\ref{lv1}.

\begin{figure}[htb]
	\centering
	\ifonecol
		\includegraphics[scale=1.0]{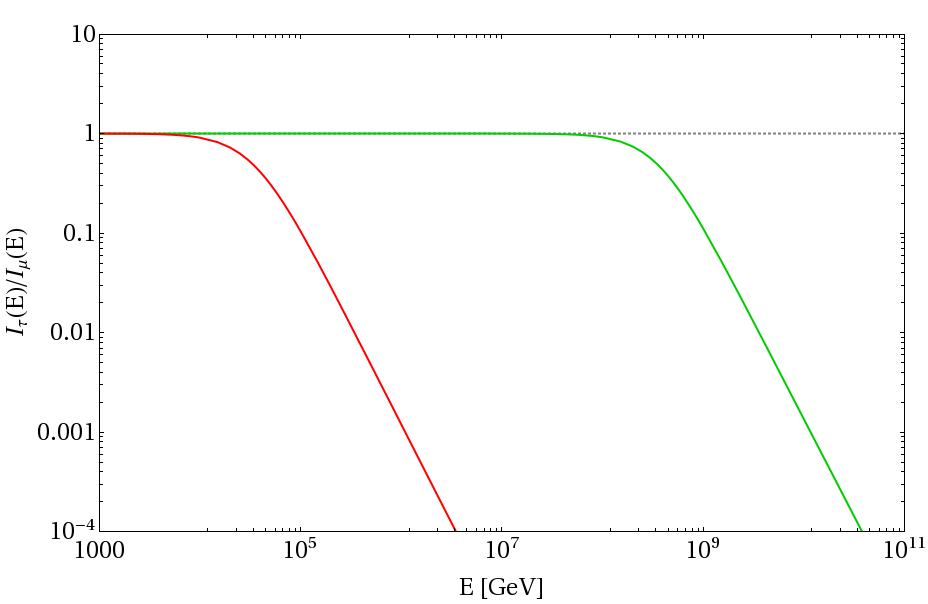}
	\else
		\includegraphics[scale=0.27]{ratio_LV}
	\fi
	\caption{\label{lv2}\textbf{Deviation of the $ I_\tau(E) / I_\mu(E) $ ratio} from that expected due to standard physics (gray dotted line) for the cases of Lorentz-violation with $ a = 10^{-26} $ GeV (red curve) and $ a = 10^{-30} $ GeV (green curve).}
\end{figure}

Fig.~\ref{lv1} shows the strong depletion in $\nu_\tau$ fluxes and corresponding enhancement of $\nu_\mu$ flux that results from this. It depicts the range over which one can test for LV and the distinctive breaking of the otherwise prevalent $\nu_\mu-\nu_\tau$ symmetry that ensues as a consequence. We find that values of $a$ from about $a=10^{-26}$ GeV to $a=10^{-30}$ GeV manifest themselves via non-observability of the double-bang, lollipop and earth-skimming events that characterize the presence of $\nu_\tau$ \cite{Learned:1994wg, Athar:2000rx,Alvarez1999,Fargion:2000iz}. We have checked that a similar depletion would occur in the cosmogenic (or GZK) neutrino fluxes even if they are produced relatively close by. Fig.~\ref{lv2} shows the strong deviation of the ratio $ I_\tau(E):I_\mu(E) $ of fluxes from the $ 1:1 $ expected due to exact $ \nu_\mu - \nu_\tau $ symmetry in standard physics.

\section{Incomplete Neutrino Decay and Neutrino Lifetimes} 

Radiative two-body decays of neutrinos are tightly constrained, as are decays to three neutrinos whereas the bounds on invisible neutrino decays to a neutrino or anti-neutrino plus a scalar or pseudo-scalar boson are relatively weak\footnote{It has been claimed in \cite{ll:043512} and \cite{PhysRevD.72.103514} that CMB data constrain such couplings strongly since they provide some evidence for neutrino free-streaming. These bounds depend on the number of species that free-stream, as pointed out in \cite{bell:063523}. Present data do not dis-allow coupled neutrino species, as the detailed study in \cite{Friedland:2007vv} shows, and further assesment of the bounds must await more data} \cite{Amsler:2008zzb,Pakvasa:1999ta,Bilenky:1999dn}. In the following we consider invisible decay modes for the heavy neutrinos while keeping the lightest stable. Neutrinos of mass $m_i$, rest-frame lifetime $\tau_i$, energy $E$ propagating over a distance $L$ will, due to decay, undergo a flux depletion given (in natural units with $c=1$) by $ \exp\left(-\frac{L}{E}\times \frac{m_i}{\tau_i} \right) $. This enters the oscillation probability and introduces a dependence on the lifetime and the energy that, in general, manifests itself in the spectrum of each flavour. The present bounds on the ratio $\tau_i/m_i$ depend on the magnitude of the (average) $L/E$ for a given source, and have the approximate values $\tau_2/m_2$ $\geq 10^{-4}$ s/eV from solar neutrino data, and $\tau_3/m_3 \geq 10^{-10}$ s/eV for the normal hierarchy from atmospheric oscillations \cite{Beacom:2002cb,Bandyopadhyay:2002qg,Joshipura:2002fb}. Including the decay factor, the flux at earth becomes\footnote{In this paper, as a first approximation,  we make the simplifying assumption that secondary neutrinos produced as a result of neutrino decays are severely depleted in energy and do not make a contribution to the fluxes in the energy range considered.}
\begin{eqnarray}
\label{simple1}
\phi_{\nu_\alpha}(E) &=& \sum_{i\beta}\phi^{\rm source}_{\nu_\beta}(E)
|U_{\beta i}|^2 |U_{\alpha i}|^2 e^{-L/\tau_i(E)} 
\end{eqnarray}
We use the simplifying assumption $ \tau_2/m_2 = \tau_3/m_3 = \tau/m $ for calculations involving decay with normal hierarchy, and similarly $ \tau_1/m_1 = \tau_2/m_2 = \tau/m $ for that with an inverted hierarchy.

\begin{figure*}
	\centering
	\ifonecol
		\includegraphics[scale=0.185]{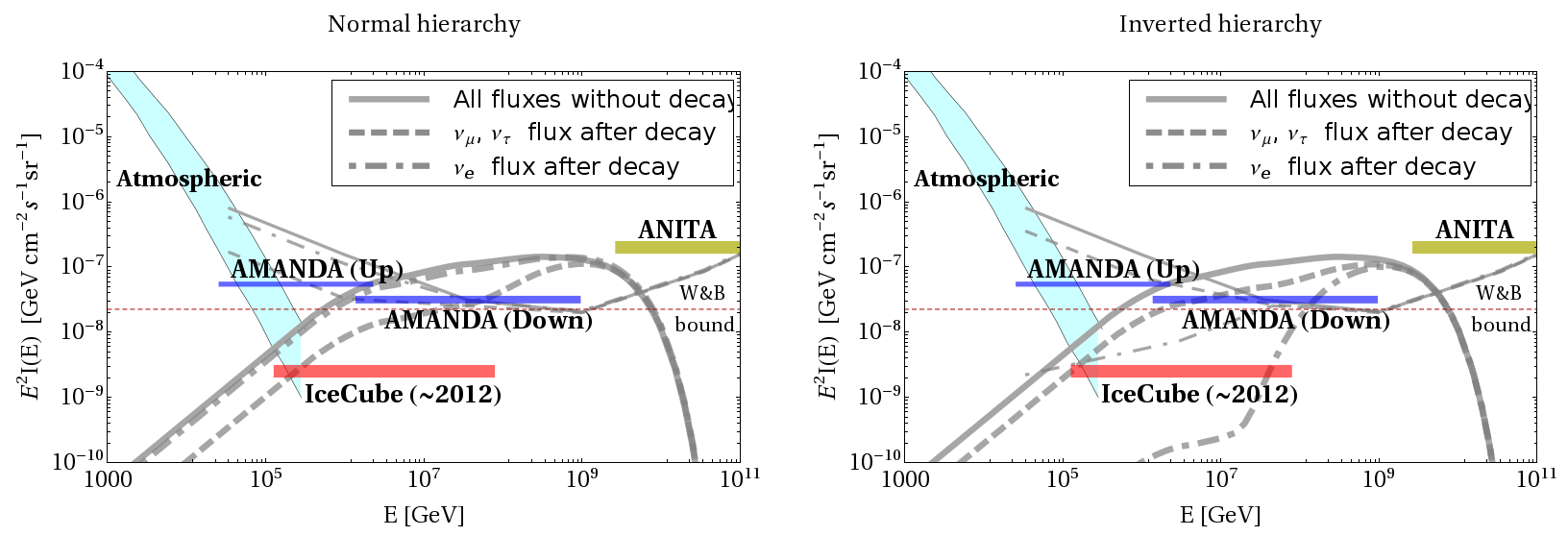}
	\else
		\includegraphics[scale=0.32]{decay_comb}
	\fi
	\caption{\label{decay}\textbf{Modification of MPR bound for incomplete decay} with normal hierarchy (left) and inverted hierarchy (right), and life-time $ \tau/m = 0.1 \;\mathrm{s/eV}$. The $\nu_\mu$ and $\nu_e$ fluxes shown are from optically thick (in thick) and optically thin sources (thinner). Similarly the gray lines indicate the $\nu_e,\; \nu_\mu,$ or  $\nu_\tau$ undistorted flux  modified only by neutrino oscillation, for both optically thick and thin sources.}
\end{figure*}

The assumption of complete decay leads to (energy independent) flux changes from the expected $\nu_e^d:\nu_\mu^d:\nu_\tau^d = 1:1:1$ to significantly altered values depending on whether the neutrino mass hierarchy is normal (\textit{i.e.} $m_3^2-m_1^2 = \Delta m_{31}^2 > 0$) or inverted (\textit{i.e.} $\Delta m_{31}^2 < 0$) as discussed in \cite{Beacom:2003nh}. From Fig.\ \ref{decay} we note that the range of energies covered by UHE AGN fluxes spans about six to seven orders of magnitude, from about $\sim 10^3$ GeV to $\sim 10^{10}$ GeV. For the ``no decay" case, the lowest energy neutrinos in this range should arrive relatively intact, {\it i.e.} $ L/E \simeq \tau/m\simeq 10^4 $ s/eV. In obtaining the last number we have assumed a generic neutrino mass of $0.05$ eV and an average $L$ of 100 Mpc. On the other hand, if there is complete decay, only the highest energy neutrinos arrive intact, and one obtains \textit{i.e.} $ L/E \simeq \tau/m\leq 10^{-3} $ s/eV. Thus, a study of the relative spectral features and differences  of flavour fluxes at earth allows us to study the unexplored range $ 10^{-3} \text{ s/eV } < \tau/m < 10^4\text{ s/eV } $ via  decays induced by lifetimes in this range (we have referred to this case as ``incomplete decay" in what follows).

To calculate the  individual flavour fluxes with decay, $\nu_\mu$ and $\nu_e$, we use a procedure similar  to that adopted for the base curves, except that we replace the flux at earth from a single source for each flavour by the modified expression in Eq.\eqref{simple1}.

\begin{figure}[htb]
	\centering
	\ifonecol
		\includegraphics[scale=0.185]{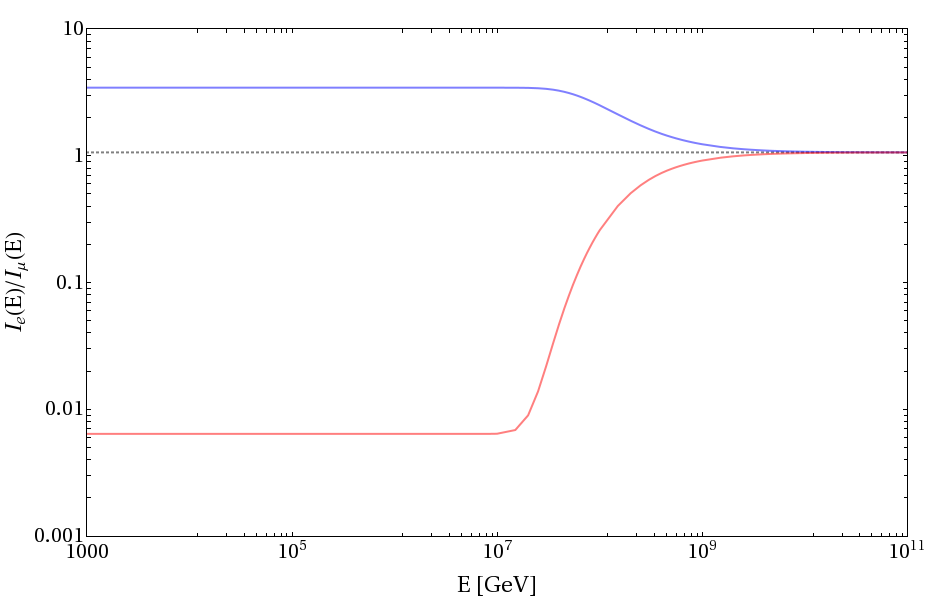}
	\else
		\includegraphics[scale=0.27]{Ratio_decay}
	\fi
	\caption{\label{fig:ratio_decay}\textbf{Deviation of the $ I_e(E) / I_\mu(E) $ ratio} from that expected due to standard physics (gray dotted line) for the cases of decay with normal hierarchy (blue curve) and inverted hierarchy (red curve). Lifetime assumed for both the heavier mass states is $ 0.1 $ s/eV.}
\end{figure}

Clearly, as is also the case for complete decays, the results are very different for the two possible  hierarchies. As is well-known, the mass eigenstate $m_1$ contains a large proportion of $\nu_e$, whereas the state $m_3$ is, to a very large extent, just an equal mixture of $\nu_\mu$ and $\nu_\tau$ with a tiny admixture of $\nu_e$. Thus incomplete decay to the lowest mass eigenstate with a normal hierarchy (\textit{i.e.} $ m_1 $) would lead to considerably more shower events than anticipated with an inverted hierarchy. Fig~\ref{decay} shows our results for these respective cases with the left panel showing the modification in the MPR bound for a normal hierarchy, assuming a representative value of $\tau/m = 0.1$ s/eV for both neutron opaque and neutron-transparent sources, while, for the same value, the right panel shows the significant distortion in the bounds that results for an inverted hierarchy. Fig.~\ref{fig:ratio_decay} shows the deviation of the ratio $ I_e(E) : I_\mu(E) $ of fluxes due to decay with the same life-times, from that expected from standard physics ($ \approx 1:1 $).

In assessing the results for decay with inverted hierarchy, which we give here, and similarly, those given in earlier work on flavour ratios \cite{Beacom:2002vi,Beacom:2003nh}, it is worth noting that this scenario is possibly constrained by the observations of $ \overline{\nu}_e $ from SN1987A \cite{PhysRevLett.58.1490}. This is because their numbers were roughly consistent with those expected, leading to a ``$ \tau/m $" for the electron flavour $ \geq 10^5 $ s/eV. If decay had occurred, and if the true hierarchy were inverted, then one would expect very few events going by the small electron flavour content of the lightest mass eigenstate $ \nu_3 $. Inspite of the fact that the total SN1987A  signal was just a handful of events, and other uncertainties associated with as yet incomplete knowledge of neutrino emission from supernovae, the results for  decay with inverted hierarchy given here should be judged with this constraint in mind.

We stress that the depletion of the $\nu_e$ flux compared to that of $\nu_\mu$ at lower energies for the decay scenario with inverted hierarchy, and its reappearance at 
higher energies is a distinctive feature. It is indicative of new physics (like incomplete decay) as opposed to spectral distortions originating in the source, or  the appearance of a new class of sources. In the latter case, a corresponding depletion and subsequent enhancement is expected in muon events. In the case of incomplete decay, by contrast, the fluxes return to the expected democratic ratio once the lifetime is such that higher energy neutrinos do not decay.

\section{Summary and Conclusions} The detection of UHE neutrinos is imminent. Several detectors will progressively sharpen their capabilities to detect neutrino flavours, beginning with ICECUBE's ability to separate muon tracks from shower events. We have shown that spectral changes in diffuse UHE neutrino fluxes of different flavours are probes of new physics entering the oscillation probability. For specificity we have used AGN sources, and calculated the changes induced in the well-known MPR bounds on both neutron-opaque and neutron-transparent sources. Our calculations can, in a straightforward manner, be repeated  for other sources, or represented in terms of the WB bounds or in terms of actual fluxes and  event rates.

We have shown that diffuse UHE fluxes are sensitive to the presence, over a relatively broad range, of tiny LV terms in the effective oscillation hamiltonian, which signal important breakdowns of pillars of presently known physics. Additionally, in the specific example of neutrino decay,  UHE detectors can probe the unexplored range $ 10^{-3} \text{ s/eV }$  $< \tau/m < 10^4 \text{ s/eV } $ via spectral differences in flavours, and thus  extend our knowledge of this important parameter. Flavour spectra in decay scenarios also differ depending on the mass hierarchy, and  provide a potent tool to probe it.

However, it must be kept in mind that the ability to distinguish between flavours in most present-day UHE detectors is as yet not firmly established. At present, detectors are capable of separating between shower events and muon-tracks which include a sum of contributions from each of the flavours rather than dinstinguishing between the individual contributions themselves. Further, the energy resolutions of present or near-future large-volume detectors like the IceCube are not very high. Consequently, deviation of the flavour spectra from the standard predictions has to be significantly large, both in magnitude and shape, for detectability in any of these detectors. Our calculations show that it is indeed possible for such large deviation to occur in the diffuse fluxes of the three flavours for suitable values of the parameters in the case of Lorentz violation as well as decay with inverted hierarchy.

Other physics scenarios to which this method may be effectively applied are the existence of pseudo-dirac states, CP violation and quantum decoherence. Although the detection of the effects discussed here will undoubtedly be challenging, given the fundamental nature of the physics to be probed, it would certainly be worthwhile.

\textbf{\textit{Acknowledgment:}} RG would like to thank Nicole Bell, Maury Goodman, Francis Halzen, Chris Quigg, Georg Raffelt, Subir Sarkar and Alexei Smirnov  for helpful discussions, and the theory groups at Fermilab and Brookhaven National Lab for hospitality while this work was in progress.

\bibliographystyle{elsarticle-num}
\bibliography{uhenu_paper_lett}{}

\end{document}